# INNOVATIVE USE OF X-RAY RADIOGRAPHY IN THE STUDY OF DAGUERREOTYPES: IDENTIFICATION OF HALLMARKS

(9232 words)

## AUTHORS

**Sara Barrio**

Sara Barrio holds a Degree in Conservation and Restoration of Cultural Heritage, specializing in Graphic Documents, from the Escuela de Conservación y Restauración de Bienes Culturales de Madrid (ESCRBC), and Supervisor of Radioactive Installations in the field of Industrial Radiography. She works as a conservator-restorer of graphic and photographic documents in private practice and as assistant conservator at the Fernán Núñez Palace in Madrid. She is currently enrolled in a Master's degree in Diagnosis of the State of Preservation of the Historical Heritage at the Pablo de Olavide University in Seville. Address: Calle Sebastián Elcano 34, Madrid 28012, Spain. Email: sarabarriorestauracion@gmail.com

**Laura Alba**

Laura Alba holds a PhD in Fine Arts from the Complutense University of Madrid. Since 2004 she has been a member of the scientific team of the Prado Museum where she is in charge of the radiographic techniques and supervisor of its Radioactive Facility. She has directed her interests towards the application of imaging techniques in the study of Art History, delving into artistic procedures and techniques. She has participated in numerous research projects and her contributions are published in exhibition catalogs as well as in specialized publications. She has previously worked in other museums and institutions such as Guggenheim Museum in Bilbao, Arte Abstracto Español Museum, Juan March Foundation and Instituto de Crédito Español Foundation. Address: Museo Nacional del Prado, Paseo del Prado s/n Madrid 28007, Spain. Email: laura.alba@museodelprado.es

**Clara M. Prieto**

Clara M. Prieto holds a Master's Degree in Heritage Conservation from the Complutense University of Madrid and a Degree in Conservation and Restoration of Graphic Art, from the Escuela Superior de Arte del Principado de Asturias (ESAPA). Since 2014 she teaches conservation-restoration at the Escuela Superior de Conservación y Restauración de Bienes Culturales (ESCRBC) in Madrid. She worked in the Atelier de Restauration et Conservation des Photographies de la Ville de Paris (ARCP) as Conservation Project Manager, as well as photograph conservator for public and private institutions such as Museo Nacional Centro de Arte Reina Sofía (MNCARS), Instituto de Patrimonio Cultural de España (IPCE) and MAPFRE Foundation. Address: Escuela Superior de Conservación y Restauración de Bienes Culturales (ESCRBC), Calle Guillermo Rolland 2, Madrid 28013, Spain. Email: claramprieto@escrbc.com

The authors report there are no competing interests to declare.

**TITLE**

Innovative use of X-ray radiography in the study of daguerreotypes: identification of hallmarks

**ABSTRACT**

X-ray radiography is an imaging technique widely used in the examination of works of art and heritage objects, however no references to its application to the study of daguerreotypes have been found.

The results obtained in this study demonstrate for the first time the usefulness of X-ray radiography for locating, identifying, and characterizing the hallmarks on the daguerreotype plates. Hallmarks are of great importance for the knowledge of these photographic objects since they provide unique information for their dating.

In this study five daguerreotypes were X-rayed using two different working procedures designed with the aim of recording the hallmarks, one by means of radiographic film and the other using a digital detector. The working procedures presented here not only enabled locating the hallmarks of all daguerreotypes studied, but also allowed its positive identification without extracting the daguerrean package or handling the plate, avoiding its exposure to damaging environmental factors.

X-ray radiography is thus revealed as an innovative and easy to integrate tool for the study of daguerreotypes, both for the characterization of hallmarks on the plate and the analysis of other elements and manufacturing procedures.

**1. INTRODUCTION**

The daguerreotype is considered the first photographic process, which gives it a high symbolic significance beyond its historical, technical and artistic value. The invention was presented to the public in 1839 by Louis-Jacques-Mandé Daguerre (1787-1851), causing a great impact on the scientific community, as well as the general public. Its use lasted until ca. 1860.

In 1895 Wilhelm Conrad Röentgen (1845-1923) discovered X-rays, which broadened the horizon of diagnostic techniques applied to scientific research and marked a new milestone in the history of imaging. Since its beginnings, X-ray radiography has been a broadly used technique in the study and characterization of a wide range of objects in the artistic field (Lang, Janet, and Middleton 2005; O'Connor and Brooks 2007), however, its use is not common in the field of photography due to the thinness of the substrates (Herrera Garrido 2022, 96). To date, no documented applications of this technique to the study of daguerreotypes have been found, possibly because X-ray radiography does not seem suitable for the study of the daguerrean image, which is the objective of most specialists' research.

However, X-rays are the foundation of other elemental analysis techniques such as X-ray fluorescence spectrometry (XRF) and X-ray diffraction (XRD) both employed for decades for the characterization of the materials present in daguerreotypes (Da Silva et al. 2010; Schulte et al. 2011; Kozachuk et al. 2018a; Sanderson, Centeno, and Stephens 2020). In recent years, the application of XRF imaging techniques enables studying the distribution of chemical elements in the daguerreotype image and generating its image when it is deteriorated (Davis, Hilton, and Vicenzi 2014; Kozachuk et al. 2018b; Stark et al. 2021).

But it should be noted that the daguerreotype image is only one of the elements that constitute the daguerreotype. In the words of Ángel Fuentes (2008, 5), "Toda fotografía es objeto complejo que requiere de al menos dos elementos para su realidad: no es posible ninguna fotografía sin un soporte y sin imagen final" ("Every photograph is a complex object that requires at least two elements for its reality: no photograph is possible without a support and without a final image.")

## 1.1. DAGUERREOTYPES

Daguerreotypes are direct photographic images, being composed of silver, mercury, and gold —if gilded—. It can appear as positive or negative depending on the viewing angle. The image support is a copper sheet (whose thickness used to be 0.4 mm) covered with a thin layer of silver (0.01 mm) polished like a mirror (Lavédrine 2007, 37).

The image is extremely fragile and vulnerable to mechanical damage and atmospheric agents, and can be damaged with the slightest touch. For this reason, it is protected by a glass, intercalating a mat of cellulosic or metallic material, and bound together with paper tape, as shown in figure 1. The daguerrean package is often secured with a flexible decorative brass "preserver" bent over the edge of it.

Finally, the daguerrean package can be inserted in a case or hinged box made of wood or other materials with a decorative finish, following the Anglo-American style; or in a simpler manner without a case, in a frame or passepartout according to the European style.

The daguerreotype plate seal prevents the action of oxygen and environmental contaminants on the silver, preventing its oxidation and sulfidation. Loss or breakage of this seal usually produces concentric interferential colored tarnish films that appear on the edges of the plate and can even cover the image of those plates without protection (Barger and White 2000, 162). Likewise, mounting systems act as a first line of defense against physical and chemical deterioration agents, slowing their advance towards the daguerreotype image (Prieto 2017). For this reason, it is essential to preserve the integrity of the original seals and housing systems, unless the conservation conditions of any of their elements make a restorative intervention necessary to ensure the stability of the artifact.

From an art-historical perspective, most of these daguerreotype portraits are anonymous and lack any visible information other than merely iconic. As Grant B. Romer (2021) points out, without identification of the subject depicted or the author of the image, these early photographic records are often undervalued and deprived of the attention they deserve.

One way to contribute to enhancing their value would be to research the aesthetic and technical information that can be obtained from the different daguerreotype elements, which would allow narrowing their dating, origin, and material history. Chiesa and Gosio (2020, 5) indicate the study of hallmarks as a means to break this anonymity, since they offer relevant information on the origin and dating of the object, the maker of the plate, importers and, on some occasions, even the photographer.

Until now, the study of these hallmarks could not be carried out without direct observation of the daguerrean plate, devoid of its mounting and sealing elements, a circumstance which, as already indicated, is neither feasible nor advisable, both for deontological and conservation reasons.

## 1.2. HALLMARKS

The origin of hallmarks on daguerreotype plates is linked to the French law[1] in force that regulated the quality of silver and other precious metals at the time of daguerreotypie.

Silver plating onto copper was usually done in two ways: either mechanically, known as Sheffield plating or *cold-roll cladding*, which consisted in fusing the two metal sheets in a rolling mill; or by an electrochemical process called electroplating, in which silver is deposited on the plate using a galvanic battery, popularized by the French firm *Christofle et Cie* from 1851 on (Barger 2000, 50). There was another silver plating process by electrolytic deposition called galvanizing (or re-silvering), introduced in the early 1840s, which was not applied onto copper but to existing silver clad polished plates, in order to reuse them after incorrect exposure or to homogenize their surface and increase their photosensitivity (Robinson 2017, 137).

Silver plating was carried out by traditional goldsmith workshops, obliged by law to stamp on their plates the quality or the proportion of silver to copper (the "mass" value), a maker's mark symbol with their initials, and, usually, the word "doublé".[2] Hallmarks may consist of one or several punches, generally the mark indicating silver content was separated from the maker's one. The plates marked with numeral 40 are the most common, indicating the silver is only one-fortieth the total thickness; plates stamped 20, 30, 50, and 60 were also produced.

As for the French term "doublé", authors such as Barger (2000, 223) stress that its presence on a French plate refers to the silver being plated, but not with which method; while the Image Permanence Institute points out through its online Graphic Atlas resource that it is a Sheffield plated silver. Both agree that the scale symbol stamped on the plate means that it was electroplated.

American manufacturers adopted the use of hallmarks by imitation of the French, but not by legal requirement, so it is unusual for them to include the silver content along with their symbol or name. The geographical area of production of the plate does not always correspond to the area of use; proof of this are the French plates, highly valued and exported to the rest of Europe and the United States.

A distinction is made between positive punches, e.g., engraved on the plate; and negative punches, which are embossed in a rectangle, oval or other figures, and the picture or number raises from the background. American plates are usually marked with positive punches, while European hallmarks often contain negative punches with more elaborate designs (Chiesa and Gosio 2020, 49).

It should be noted that hallmarks are not necessarily present on all plates. They used to be stamped on the recto of whole plates, in some of the corners or nearby areas, so this information was lost when they were cut to smaller standardized formats, or to facilitate their insertion in plate-holding devices during manufacturing or in protective cases.

While it has been shown that the analysis of the clothing and hairstyle of the portrayed can be useful for dating the image (Severa 1996), the identification of makers' hallmarks is the most reliable method for this purpose, at least for establishing the "not earlier than" date (Berg 2020, 19). The protective elements of the daguerreotype can provide indicative and complementary data to be taken into account, but it is not always possible to ensure its provenance unless it conserves its original seal and does not show evidence of having been manipulated.

## 2. OBJECTIVE OF THE STUDY

The objective of this research is to substantiate the usefulness of the radiographic technique in the study of these complex and delicate objects. There are many features in daguerreotypes that can be observed in an X-ray radiography, however the first phase of this project is restricted to locating and interpreting the small hallmarks engraved on the copper plates without disassembling the daguerrean package.

Recording the hallmarks on an X-ray radiography is a challenge for several reasons. First of all, because of their small size, sometimes only 2.4 mm long, and because of the subtle thickness of the engraved motifs, sometimes difficult to see even with the naked eye. Added to this is the complexity of the object in its own structure, since all the elements of the package appear overlapped on the same image, interfering with the reading of the radiograph. In the case of hallmarks, the brass preserver protecting the daguerrean plate, the mat when decorated and the cardboard backs are particularly disturbing as they interfere to observe small details.

The possibility offered by X-ray radiography to locate and identify the hallmarks without handling the daguerrean package, even without removing it from its housing, is an added tool to the non-destructive techniques, contributing to the knowledge of these unique objects.

## 3. MATERIALS AND METHODOLOGY

The radiography procedure used can critically determine the results obtained and therefore the effectiveness of the technique. For this reason, and in view of the wide variety of industrial radiography systems on the market, two working procedures have been designed, which differ mainly in the focal size of the X-ray tube and in the image recording system. The first procedure focuses on equipment and methodologies common in the artistic field, while the second one is a less frequent methodology, but more suitable for this type of object and for the objective of the research: to locate the hallmarks and identify the elements of which it is composed.

In order to evaluate the suitability of these working procedures, thirteen daguerreotypes of diverse morphological characteristics and housing systems were radiographed, five of which were considered the most representative objects and thus selected , both for their morphological and presentation structures, and for the results obtained with the two radiographic procedures. The selected daguerreotypes belong to Spanish private collections and their authors, their exact dating and the identity of the people portrayed were unknown.

## 3.1. IDENTIFICATION OF THE DAGUERREOTYPES

**Daguerreotype no. 1** is a seated portrait of a mother with her daughter (fig. 2). Its format corresponds to one sixth-plate. The daguerrean package is composed of a brass preserver, a cover glass, a metal mat, and the plate; it preserves the original seal. It is presented protected in a wooden case covered with embossed leather, which only conserves the tray. A paper with an ink handwritten inscription indicating the maternal-filial relationship between the characters portrayed is attached to its verso.

The design of the mat has allowed a first estimation for dating the daguerreotype between late 1840s and mid 1850s, a period in which this typology was in use (Berg 2020, 33; Nolan 2021, 41). The dating of the rest of the elements according to their decoration has been carried out through the catalog developed by Nolan (2021). The decoration of the preserver corresponds to that classified as "*s-bunting*", and is dated by the author between 1850 and 1857 (115). As for the case, the leather embossing shows a medallion containing three roses, matching a piece listed between 1853 and 1856 (192). The interior of the case is also decorated with a gilt edge (reference "*gStraightBracelet*") which was in use between 1855 and 1859 (247).

All the protective elements perfectly fit each other and the object , so they are considered to be original. This, together with the fact that their dating is consistent, would place the object in the mid 1850s.

The original seal had lost its function, which facilitated access to the daguerrean plate and confirmed the existence of a hallmark located in the lower left corner. This circumstance was especially relevant because it allowed studying the hallmark with visible light as well as X-ray radiography, therefore allowing to compare both images, the visible and the one registered in the radiographic image, in order rigorously and objectively assess the reliability of the technical results obtained.

The punch is made up of two sections of letters, a curved upper section and a straight lower section parallel to the edge of the support. Its dimensions are 3.5 x 11 mm, although it appears incomplete.

It is also noted that the daguerrean plate presents four symmetrical tabs, two on each side, which bend towards the verso of the plate. They are 2.5 mm long and rectangular in section. These tabs' function was to hold the plate onto a fixing device during the polishing process.

**Daguerreotype no. 2** is a half-length portrait of a man (fig. 3). It has a format corresponding to one sixth-plate. The daguerreotype package consists of a brass preserver, a cover glass, a metal mat, and the plate, retaining its original seal that shows no signs of having been tampered. The object is mounted in a leather-covered wooden case the structure of which remains intact .

As in the previous case study, the Nolan catalog (2021) has been used to date the protective elements based on their decoration. The mat is an oval-format window with extensive use from 1840 to 1865; it features a grape leaves decoration that the author dates between 1854 and 1856 (72). As for the preserver, its decoration shares remarkable similarities with a specimen ("*Nolan s-ivy*") dated between 1852 and 1855 (116). Looking at the decoration of the case, the embossing of the leather covering correlates to piece #211 (208), previously recorded by Rinhart and Rinhart (1969), and dated by both ca. 1853. The interior of the case has a decorative gilt edge ("*Nolan gJacobsLadder*"), Nolan notes between 1854 and 1855 (247). This data places the daguerreotype in the mid 1850s.

**Daguerreotype no. 3** is a half-length portrait of a man (fig. 4). Its format corresponds to one quarter-plate. The daguerrean package is composed of a metal mat, the plate and a cover glass which thickness indicates that it is of that period. The assembly is sealed with a modern paper tape that maintains the tightness. The mat has a double elliptical shape, granulated texture and lacks decoration, this simplicity being a generalized trend in the 1840s that will evolve towards a progressive complexity and sophistication from 1850 onwards. According to Nolan (2021, 14), daguerreotypes with this type of mat opening are usually of French origin and this particular model dates between 1847 and 1853 (45). However, it is not possible to be sure that the protective element is original because the daguerrean package shows signs of manipulation.

**Daguerreotype no. 4** is a seated portrait of a woman (fig. 5). Its dimensions correlate to one sixth-plate. It conserves only the daguerrean package, consisting of a brass preserver, a cover glass, a metal mat, and the plate. It is hermetically sealed with a modern paper tape, evidence that brings into question the originality of the protective elements. Dating according to the decoration present on the mat, based on a line of indented dots at a short distance from the cut out, varies depending on the bibliographic reference source used : according to Berg (2020, 34) it was introduced in the mid to late 1840s; however, Nolan (2021, 46) indicates a later use, between 1854 and 1859. The preserver features a plant decoration that Nolan dates between 1862 and 1865, already at the end of the period of daguerreotype use, which not only reflects a discrepancy between the dates of the protective elements but also indicates that this element comes from another cased photographic artifact of later use (an ambrotype or a ferrotype).

**Daguerreotype no. 5** is a half-length portrait of a man (fig. 6). It is presented in European format on a thick paper passepartout with an oval window and golden bevel, and a cardboard back. This mat window design was introduced in the late 1840s (Barcella 2009, 7). The thickness of the glass, marked on the paper, is regular and measures 2 mm, inferring that it is not an original industrial glass. A modern-looking black paper that was probably added when the primary glass was replaced seals the set. The plate shows tarnish lines following the shape of the protective element, suggesting it could be original. A paper label containing an ink handwritten inscription (not legible), is attached to the verso. The overall dimensions of 12.5 x 15 cm suggest that plate format corresponds to a quarter (8.2 x 10.7 cm).

## 3.2. RADIOGRAPHIC STUDY

### 3.2.1. Preliminary Considerations

Radiography is an imaging technique based on the ability of X-rays to pass through materials. Although outside the human visible spectrum, this electromagnetic radiation travels in a straight line, and is capable of generating images by impressing photographic films and fluorescent materials. The image obtained is determined by the energy of the incident radiation and by the ability of the object to attenuate the pass of the radiation and, therefore, by the composition, density, and thickness of the materials that constitute it. In the case of daguerreotypes, it is a mixed object with a complex structure materials of which have different characteristics and therefore different radiation attenuation coefficients. All these materials: copper, silver, brass, wood, leather, glass, or fabric can be X-rayed, although their recording is determined by the exposure conditions and the materials overlapped on them.

Daguerreotypes are particularly delicate objects that require preventive conservation measures to ensure their stability. While atmospheric pollutants and humidity constitute the greatest risk to these pieces, some specialists have also included light among the factors to be controlled. Thus, if they were sensitive to radiation in the visible spectrum, it could be considered that X-rays, which are more energetic, could also damage them. This last aspect is particularly controversial, since there are specialists such as Lavédrine who do not consider it a risk factor in non-colored daguerreotypes, but at the same time advise "to avoid overly intense light sources during display" without explaining why (Lavédrine 2009, 29-30). In any case, before using the radiographic technique, as with any technique that uses X-rays, it seems necessary to assess whether the daguerreotype, including the daguerreotype image, the package, and the case, can be altered when radiographed.

In this regard, several investigations have been published in recent years studying the mechanisms of occurrence of silver sulfides and chlorides that trigger alterations in the daguerreotype image. Among the factors that cause tarnishing by the presence of sulfides environmental aspects such as "relative humidity, temperature, the amount of S in the atmosphere, and the state of the Ag surface" are cited (Kozachuk 2019, 1684). Neither visible nor ultraviolet spectrum radiation are considered risk factors (Bossi, Iddings, and Wheeler 2002, 234). Regarding the chloride formation, some authors have proposed that their formation is related to redeposited silver processes from AgCl crystals, in which case it is proposed that visible light favors the process and that it would be irreversible (Centeno, Meller, and Kennedy 2008). On the contrary, other authors in recent studies, claim that the mechanism is different and spontaneous, and that it is due to the so-called Ostwald ripening process, which causes the recrystallization and coarsening of AgCl crystals (Robinson 2015). In this case the process would be reversible by elimination of chlorides and visible light would not be a determining factor, since haze deterioration has been described in daguerreotypes preserved in the dark and does not develop in others exposed for long periods to daylight.

In addition, as mentioned above, daguerreotypes have been studied for several decades with elemental analysis techniques using X-rays, and no alteration caused by the use of these techniques has been described. Moreover, it should be taken into account that the duration of the exposure of the daguerreotype image to ionizing radiation to obtain a radiograph is extremely short, in some cases milliseconds.

Thus, X-ray radiography in the study of daguerreotypes can be considered a non-destructive technique, as it does not damage the materials it passes through or trigger any harmful chemical reaction, and non-invasive since it is an imaging technique that does not require sampling.

### 3.2.2. Radiographic Procedures

**Procedure 1:** Radiographic film and milifocus X-ray tube

A radiographic procedure was designed with the objective of reproducing the traditional work methodology, using a milifocus tube and conventional radiographic film, which is still used in many museums and art institutions. A portable constant potential equipment of the Seifert brand, model ERESCO MF4, with a focal size of 1mm, was used. A 2 mm thick aluminum filter was placed at the exit of the tube to optimize beam energy eliminating those useless low energies. The optimum exposure conditions were 90 kV voltage, 6 mA intensity and a time of 7 min, at a focus-film distance of 200 cm. Fujifilm 50+Pb radiographic film with 0.27 mm thick lead reinforcing screens on both sides was used to record the image in order to reduce scattered radiation and improve detail definition. The front view of the object was recorded with the back of the daguerreotype facing the film. Once the exposure was made, the radiographic film was developed in a Fujifilm brand industrial developer model FNDX 3.0b for 8 minutes and at a temperature of 26ºC. The radiographic plates were visualized in a KOWOLUX model LX60 industrial negatoscope with the aid of a 10x magnifying glass.

In order to obtain a record that would allow digital processing, the daguerreotype plates studied were digitalized with two systems: photographed in the negatoscope with a Canon Eos RP digital camera with a 24-105 mm lens, and digitized with a specific scanner for radiographic film from Array Corporation. All digitalized images in TIFF and grayscale format were displayed on an Eizo RadiForce RX430 monitor.

The digital images were processed with Adobe Photoshop CS5 and the same enhancement procedure with specific tools (layers, levels, and focus mask filter) and identical in all cases in order to compare the results obtained.

**Procedure 2**: digital detector and microfocus X-ray tube

The second procedure was designed with the intention of obtaining an optimal resolution that would allow quality reproduction of the small details of the hallmarks. For this purpose, a Hamamatsu X-ray source model L9181-02 with a minimum microfocus of 5 μm and a maximum energy of 130 kV was used (Fig. 7). The advantage of these tubes is that the micrometer-sized focal spot minimize the penumbra effect and the images can be highly magnified with negligible loss of sharpness. To increase the effective energy, the radiation was filtered at the tube output with 1 mm thick aluminum plate. Images were recorded using an amorphous silicon flat panel digital detector (Varex, model XRPad2 4336 HWCi) with a pixel size of 100 μm.

The focus-detector distance was 100 cm in all exposures, while the focus-object distances varied between 3 and 100 cm with the intention of amplifying the signal. Typical magnification factor was 20 which corresponds with a focus object distance of 5cm and yields an effective resolution of 5 microns. The X-ray exposure parameters were typically set at values of 130 kV energy, 50 μA intensity . Detector exposure time was set at a value of 2800 ms with a digital gain of 1 and 32 averages in order to improve the image quality and optimize SNR (Signal-to-noise ratio).

The images generated by the digital detector were displayed on the aforementioned Eizo monitor and processed in Photoshop following the same procedure described for the digitalized radiographic film.

# 4. RESULTS AND DISCUSSION

## 4.1. RADIOGRAPHED PIECES

**Daguerreotype no. 1.** The X-ray radiographies obtained by the two radiographic procedures reveal the existence of a hallmark in the lower left corner, as well as four cut corners in a regular shape (fig. 8). It can be seen that the letters in the hallmark have greater radiographic density than the surrounding surface, indicating that they are incised on the plate and that it is, therefore, a positive punch.

Regardind the radiograph obtained with Procedure 1, it was corroborated that the dimensions of the hallmark, measured directly on the daguerreotype plate and on the radiographic film, are identical: 3.5 x 11 mm (see 4.2 Procedures).

In the X-ray radiography it is not possible to recognize all the letters forming the hallmark due to a lack of homogeneity in the pressure exerted at the time of stamping, a feature that is difficult to see even with the naked eye. However, this has not prevented its positive identification with the one referred to in the catalog of Chiesa and Gosio (2020, 100) with the code "SCOVILL (Rinhart n.46c)" (fig. 9).

The hallmark belongs to the last production period of the Scovill company in the 1850s. The brothers J. M. Lamsom and William H. Scovill were the first and most important manufacturers of daguerreotype plates in the United States. The company, founded in Waterbury (Connecticut) in 1802, adapted its plating business to the new photographic market. Under the name *Scovill Manufacturing Company*, it began marking daguerreotype plates as "SCOVILL MFG-EXTRA" beginning in 1850 (Chiesa and Gosio 2020, 99). This new plate was lighter and more ductile, and with a very smooth silver finish (Rinhart and Rinhart 1981, 168). According to Barger and White (2000, 51), the Scovill Extra was made by electroplating, following the successful silver plating process introduced by *Christofle et Cie* from 1851, a fact that would explain the improvement in its quality.

Identifying the manufacturer of the daguerrean plate primarily dates the artifact between 1851 and 1860. Taking into account the information provided by the protective elements, preserving the original seal and showing no signs of manipulation, it is concluded that daguerreotype no. 1 was made between 1851 and ca. 1855, and that the silvering of the plate was done by electroplating.

Four symmetrical tabs, two on each side, which had been previously seen in the daguerrean plate, are recorded in the radiography with lower radiographic density, although they are observed with better definition and dynamic range in the image obtained with Procedure 2. These notches must have been made when the plate was fixed in one of the devices used during the manufacturing processes, such as buffing, polishing, or gilding. Some of these devices were patented, although in this case it has not possible to relate them to any specific one. It should be noted that the deformations they generate can be detected by the radiographic technique. Also, the X-ray radiography of this daguerreotype shows information about its structure and the elements of its case.

**Daguerreotype no. 2.** As it can be seen in figure 10, the radiographies obtained by the two radiographic procedures show the existence of a hallmark located in the upper right corner. In spite of the fact that the decoration on the preserver and the mat makes the reading of the hallmark notably difficult, both procedures reveal that it is a negative punch of rectangular shape whose background is registered in the radiograph with greater radiographic density than in the surrounding areas. This is because it is slightly recessed on the daguerrean plate, so that the symbol, letters and numbers it contains are embossed with respect to the background. Its dimensions are 3 x 6 mm.

It has only been possible to recognize all the elements that compose the hallmark in the image obtained by Procedure 2 due to its better definition and dynamic range. It has been identified as the one registered with the code "HBEH40 (Rinhart n.20)" in the catalog of Chiesa and Gosio (2020, 84).

This is a negative hallmark with initials "H. B." dotted, eagle, and 40 silver mass, inscribed in rectangle with clipped corners. The eagle, its wings folded and in frontal position, crowned with a star and holding a globe in its claws. According to Rinhart and Rinhart (1981) it is probably the most popular French plate.

This punch is associated with the French manufacturer Henry Beaud, whose business was located in Paris. Its presence on the plate indicates that Daguerreotype no. 2 was manufactured between ca. 1850 and ca. 1858, the period covered by their production of daguerreotype plates (Chiesa and Gosio 2020, 84).

Taking into account the information provided by the decoration of the protective elements of the object and, given that it retains its original seal and shows no evidence of tampering, it has been dated between 1853 and 1854.

Likewise, the general X-ray radiography of the piece shows that the daguerrean plate has four cut corners cut in a regular shape and that it has a recess in the central area of each side (fig. 10a). These distortions are located 2 cm from the upper edge of the plate and have a length of 3.6 cm.

In addition, the radiography obtained by Procedure 2 allows to detect that the plate has the perimeter edges slightly bent towards the back of the plate. This feature is recorded on the X-ray radiographies as a thin line of lower radiographic density located at the edge of each side of the plate.

All of these distortions are possibly associated with one or more plate fixation devices. As indicated by Rinhart and Rinhart (1981, 177-78), the four cut corners and the perimeter edges slightly bent towards the reverse are characteristic evidence of the use of one of the most popular plate holders, that of the daguerreotypist Samuel Beck, patented in 1850, a date consistent with the dating of the object. In this case, a second device could have generated the deformations on the sides.

**Daguerreotype no. 3.** The radiographies obtained are easy to read due to the simplified structure of the object and the absence of decoration on the mat (fig. 12). The two radiographic procedures confirm the presence of a hallmark in the upper right corner of the daguerrean plate oriented in the opposite direction to the reading of the image, and also show that its corners have been irregularly cut.

As it can be seen in the X-ray radiographies, the hallmark is made up of three adjacent punches, two of them of rectangular shape and a third one barely perceptible as it is interrupted by the corner cut. The background of the rectangles is registered in the radiograph with greater radiographic density than in the surrounding areas, so they are negative punches. The one stamped with numeral 40 indicates the silver content and measures 3 x 3.5 mm. The second one is the maker's mark and it shows his symbol and his initials, in this case an Agnus Dei flanked by two moons, the letters "J", and "P" on the lower border, as well as the word "DOUBLÉ" on the upper border; it measures 3.5 x 4 mm.

The hallmark has been identified as the one registered under the code "AJPD40 (Rinhart n.29)" in the Chiesa and Gosio catalog (2020, 63) (fig. 13). It belongs to Jean-Baptiste Pillioud (JP), a French silversmith whose production began in the early 1800s until 1846, when he sold his plating factory, located in Paris, to Alexis Gaudin[3] (Pellerin 1997, 73). Gaudin's punch practically kept the appearance of the J.P. Pillioud one, so they are easy to mistake. The initials "JP" were replaced by "A. Gaudin" and the font of the silver content mark was changed. Plates with both punches were widely used in France and exported to the United States, where they became popular (Rinhart and Rinhart 1981).

In the case of this hallmark, it has been considered that the term "Doublé " does refer to silver plating on copper by mechanical means. Being an early French plate, the use of electrolysis for this purpose is ruled out. Furthermore, in Marc Antoine Gaudin's correspondence of 1846 there are specific mentions of the rolling mills acquired by his brother Alexis when he purchased the business from J.P. Pillioud (Pellerin 1997, 75).

The identification of the hallmark has made it possible to date the daguerreotype between 1840 and 1846, overlapping with the early years of the daguerreotype process, as well as attesting its French manufacture. This data is consistent with the simple and undecorated shape of the mat, a typical trend of this period, as well as with the absence of the preserver, use of which was not introduced until 1843 according to Berg (2020, 34) or 1847 according to Rinhart and Rinhart (1981, 423). Based on these data, it is considered likely that the mat is original and that the object did not have a preserver in origin. Also, the hallmark has provided information on the silver plating process used.

**Daguerreotype no. 4.** Both radiographic procedures make it possible to locate a hallmark in the lower right corner arranged vertically and to discover that the daguerrean plate has regularly cut edges (fig. 14). As in Daguerreotype no. 3, it is a negative punch and, although it is also slightly interrupted by the corner cut, it has been found to measure 2 x 8 mm.

The hallmark consists of a blunted rectangle enclosing the initials "H.B.H.", a side face eagle which, in this case, is partially differentiated in the radiograph, and the number 40 indicating the silver content. It belongs to the Holmes, Booth, and Haydens Company of Waterbury (Connecticut), founded in 1853 by Israel Holmes, John C. Both, and the Haydem brothers ─Hiram Washington, Henry Hubbard, and Henry Hotckkiss─ and engaged in the manufacture of a very broad line of photographic products. They produced daguerrean plates between 1856 and 1861 (Chiesa and Gosio 2020, 84).

From 1856 to 1857 the company was assisted by August Brassart, the Parisian platemaker who made the world's first successful daguerreotype plate for Daguerre (Rinhart and Rinhart 1981, 155). This could be related to the fact that the hallmark includes information on the silver content, something that, as already indicated in section 1.2, was not usual in the plates manufactured in the United States.

The identification of the hallmark has allowed dating the daguerreotype between 1856 and 1861, being in this case the only reliable means to do so. Although according to the consulted bibliography these dates match those of the mat, the originality of this element cannot be guaranteed, and the preserver comes from another cased photographic artifact of later use than the daguerreotype.

**Daguerreotype no. 5**. The general X-ray radiography shows the fastening and securing system of the access door of the mount by means of four metal pins, discernible due to their radiopacity (fig. 16). The presence of numerous particles with a high radiation attenuation coefficient and irregular size distributed over the entire surface of the object is striking. It is likely that these are mineral fillers present in the composition of the cardboard (O'Connor and Brooks 2007, 122) or impurities associated with the low quality of this support.

As for the daguerrean plate, the radiographic image reveals that it is slightly tilted and has all four corners cut in a regular shape. A semicircle with a diameter of less than 2 mm is detected in the upper right corner (fig. 16b). It could be a hole ─in this case sectioned in half─ which, according to Chiesa and Gosio (2020, 148), was made in one of the corners to suspend the plate in a galvanic bath during the re-silvering treatment. An electroplating treatment could have been performed instead, since the procedure is similar in both cases.

Also, a hallmark has been detected on the right edge of the plate, near that corner (fig. 16c). This is registered in the radiographic image as a lozenge-shaped area that has greater radiographic density than the surrounding area. As it can be seen in the radiograph, its stamping has generated a slight deformation in the plate.

The readability of the elements of this hallmark is compromised due to its strikingly small size and the aforementioned presence of radiopaque particles in the composition of the board. Despite this, Procedure 2 has made it possible to identify it as the one registered in the Chiesa and Gosio catalog (2020, 67) under the code "BALCDD" (fig. 17).

It is a negative hallmark in the form of a lozenge, as already mentioned , in which two stars are inscribed at the top; in the center, a balance flanked by the initials "C C" and a flower, and a bee below. Chiesa and Gosio (2020) indicate that its length is 2.4 mm, which shows its great complexity, and date its use in years around 1844. Although these authors identify features in common with Christofle brand, they do not confirm its authorship, but on page 104 of their catalog they indicate its presence on a plate next to a punch with the name "CHRISTOFLE".

On the other hand, Nolan (2020) classifies it as a rare variant of this one. Caroline Radenac, current Heritage manager of *Maison Christofle*, confirms to the authors in a personal communication (December 6, 2022) that this hallmark corresponds to the symbol used by Christofle for silver-plated metal from 1845 on and that it contains the same identifying elements: the balance, the letters "C C" referring to the initials Charles Christofle, the stars, and a bee. Radenac points out that *Christofle et Cie* manufactured plates for daguerreotypes between 1850 and 1855, most of them being produced between 1850 and 1851, dates that differ with the one provided by Chiesa and Gosio.

This hypothesis about the authorship of the mark is reinforced by the presence of the previously mentioned hole, consistent with the fact that *Christofle et Cie* manufactured the plates by electroplating process.

Charles Christofle (1805-1863) was a French goldsmith and silversmith. In 1830 he founded *Christofle et Cie* in Paris, one of the most important and innovative goldsmith businesses of its time. In 1842 he bought from the English brothers George R. and Richard Elkington the patent for electroplating and gilding by electrolysis for commercial exploitation in France, becoming the first goldsmith to apply the electroplating process on an industrial scale (De Ferrière Le Vayer 1994).

The excellent quality of the electroplating of his famous "scale plates" meant that they were widely exported to England and the United States, and other companies, such as the aforementioned Scovill, adopted this method of plate manufacture (Barger 2000, 51).

Based on the identification of the hallmark and taking into account the data collected on its mounting system, the dating of the daguerreotype has been estimated between 1850 and 1855, and it is considered highly likely that the plate was electroplated.

## 4.2. RADIOGRAPHIC PROCEDURES

The hallmarks could be located with the two radiographic procedures in all the daguerreotypes studied, even in those more complex due to the small size of the mark or the overlapping of elements. In terms of reading and interpretation, as it would be expected, Procedure 2 has been more effective, although conventional film radiography has also totally or partially recorded the marks of daguerreotypes no. 1, no. 2, no. 3, and no. 4, which punches measuring between 3.5 and 13 mm in length.

The quality of the digital image, determined by its resolution, bit depth, and dynamic range, varies according to the process used to obtain it. This quality has determined the ability to read the hallmarks. For this reason, the mark of Daguerreotype no. 5, with a length of 2.4 mm, could only be identified in the digital image obtained with Procedure 2. The minimum size and thickness of the small symbols and letters of this hallmark means that not even the digitized radiographic film can register it.

Thus, Procedure 2, with microfocus tube and digital detector, allows obtaining better quality digital radiographies thanks to its bit depth and wide dynamic range. The resolution when the daguerreotype is placed in contact with the detector is 260 dpi at 1:1 scale, lower than with the digitized film of 513 dpi at 1:1 scale. However, the use of the microfocus tube and the possibility of magnifying the details, bringing the daguerreotype closer to the X-ray focus, makes it possible to increase the resolution of the radiography. Thus, in the tests described, values of 1200 dpi at a scale of 1:1 have been obtained, making it possible to magnify small details without losing sharpness.

The high quality of these radiographs has made the complete reading possible of all the hallmarks.

As for Procedure 1, the radiographic film was viewed on the X-ray viewer and all the hallmarks were located directly, although they were observed in greater detail using a 10x magnifying glass. This is a simple and useful procedure for reading the symbols, numbers, and letters of the hallmarks. However, it has been found that magnifying glasses with magnifications higher than 30x distort the image by magnifying the film grain and noise, making it difficult to see the details.

On the other hand, the radiographic film photographed in the negatoscope has obtained general digital images of the daguerreotype of poorer quality than when digitized with the digitizer. However, in the case of the details of the hallmarks, the possibility of using a macro lens on the camera has sometimes made the details of the marks best seen when photographed.

Despite its limited quality in attesting details, Procedure 1 shows two benefits to be taken into consideration. On the one hand, it uses cheaper equipment that is frequently used in museums and art centers, so its application in the study of daguerreotypes can be immediate in many institutions. On the other hand, the flat format of the daguerreotype, which allows good contact with the film, as well as the wide focus-object distance used, have minimized the effect of geometric distortion, and the X-ray radiographic images obtained can be considered at a 1:1 scale with respect to the originals. It has been confirmed in Daguerreotype no. 1, hallmark of which was observed and measured on the daguerrean plate itself, that real measurements can be made both on the radiographic film with a thread counter and on the digitized plate by means of image processing programs. Knowing the dimensions of the hallmark is a relevant fact when typifying and interpreting it. This fact is an important advantage compared to digital radiography, since it allows typifying the marks and making measurements in the technical document in a precise and reliable way.

Therefore, as an improvement to Procedure 2 and with the aim of making real measurements on the digital image obtained, it is proposed to use a radiopaque material of known dimensions (e.g. a lead wire of 5 cm in length). This element will be placed before taking the radiography, next to the daguerreotype and on the same geometrical plane , so that the recorded digital image can be reliably resized and scaled to its real size.

## CONCLUSIONS

The application of X-ray radiography in the study of daguerreotypes is unprecedented, although this research has proven that its use is relevant to locate and identify hallmarks in sealed and housed plates. The data is obtained non-invasively, which guarantees the integrity of the daguerreotype, an extraordinarily vulnerable object to both physical and chemical deterioration. Handling is minimal and analysis does not even require removal of the daguerrean package from its housing, thus ensuring the preservation of this unique and delicate photographic artifact.

Radiography is also an imaging technique that can be quickly incorporated into the methodology for the study of daguerreotypes as it is an equipment often used  in the artistic field.

The  radiographic study of the five selected objects has confirmed its value as a technique for locating and reading the small hallmarks engraved on the daguerreotype plates. X-ray radiography has allowed their typification in a precise, objective and reliable way, something that is sometimes difficult to achieve through direct study due to the small size and complexity of some of them. The possibility of linking the symbols, letters, and numbers of these punches with the data documented in the reference bibliography has helped stablishing the chronology and origin of the plates, as well as the method of plating, making it possible to contextualize these objects in a more precise space and time, and to contribute to enhance their value.

This article focuses on the study of the hallmarks, but beyond the proposed objective, its effectiveness has been substantiated for observing other characteristics related to the manufacturing process, both of the plate and of the daguerrean package, and even of the housing systems. This work has broken new ground for further research. In future phases of the study it is intended to identify features detectable by the radiographic technique, such as the traces left by the plate-holding devices on the plate ─bent corners and edges, indentations, etc.─ or the structure of the cases. This data will provide useful information for dating the daguerreotypes and would allow to deepen and expand our knowledge of these early and exquisite photographic portraits.

**ACKNOWLEDGMENTS**

The authors would like to thank Dr. Eusebio Solórzano from Novadep Scientific Instruments ([www.novadep.com](www.novadep.com)) for his collaboration and generosity in taking the digital radiographs at the company facilities. To Óscar Solé from Ipunto ([www.ipend.es](www.ipend.es)) and to Centro de Formación en Tecnologías del Frío y la Climatización de Madrid for their involvement in the execution of the radiographic film tests. To Josina LLera and Ascensión Fernández for their graphic contributions, and to Lucía Lorenzo and Sonia Tortajada for their suggestions to the text.

**NOTES**

1. Law of the 19th Brumaire, An VI (November 9, 1797) (Arminjon, Beaupuis, and Bilimoff 1994).

2. Alexis Gaudin (1816-1894) was an important dealer in photographic articles and manufacturer of daguerreotype plates. Together with his brothers Marc Antoine and Charles, he directed the famous journal *La Lumière* (1851-1867), the first medium devoted to scientific advances in photography.

3. The French term "doublé" originally referred to the plating of silver on copper by soldering and mechanical means, a technique widely used in France from 1768 (Bouilhet 1945, n.p.). The "doublé" maker's punch was used until the end of the 19th century, some time after the electrolysis silver plating technique ─electroplating─ became widespread (Arminjon, Beaupuis, and Bilimoff 1994, 22).

**FIGURE CAPTIONS**

Fig. 1. Outline of the structure of a cased daguerreotype (Anglo-American style).

Fig. 2. Daguerreotype no. 1. *Portrait of a mother and her daughter.* (a) Daguerreotype on its tray (b) Daguerrean plate, hallmark on bottom left corner. Unidentified photographer. One-sixth plate: 7 x 8.3 cm. Private collection. © Carlos Vacas.

Fig 3. Daguerreotype no. 2. *Portrait of a man.* Unidentified photographer. One-sixth plate: 7 x 8.3 cm. Private collection. © Carlos Vacas.

Fig.4. Daguerreotype no. 3. *Portrait of a man.* Unidentified photographer. One-fourth plate: 8.2 x 10.7 cm. Private collection. © Carlos Vacas.

Fig. 5. Daguerreotype no. 4. *Portrait of a woman.* Unidentified photographer. One-sixth plate: 7 x 8.3 cm. Private collection. © Carlos Vacas.

Fig. 6. Daguerreotype no. 5. *Portrait of a man.* Unidentified photographer. One-quarter plate: 8.2 x 10.7 cm. Private collection. © Carlos Vacas.

Fig. 7. X-ray process of the daguerreotypes with Procedure 2. The X-ray tube with microfocus is on the left, while on the right can be seen Daguerreotype no. 5 in a vertical position and placed in front of the digital detector. Novadep Scientific Instruments facilities.

Fig. 8. (a) General X-ray radiography of Daguerreotype no. 1. (b) Detail of the hallmark with Procedure 1. (c) Detail of the hallmark with Procedure 2.

Fig. 9. Reproduction of the hallmark "SCOVILL (Rinhart n.46c)" present in Daguerreotype no. 1. Source: Chiesa and Gosio (2020). 3.5 x 13 mm.

Fig. 10. (a) General X-ray radiography of Daguerreotype no. 2. (b) Detail of the hallmark with Procedure 2. (c) Detail of the hallmark with Procedure 1.

Fig. 11. Reproduction of the hallmark "HBE40I (Rinhart n.19)" present in Daguerreotype no. 2. Source: Chiesa and Gosio (2020). 3 x 6 mm.

Fig. 12. (a) General X-ray radiography of Daguerreotype no. 3. (b) Detail of the hallmark with Procedure 2. (c) Detail of the hallmark with Procedure 1.

Fig. 13. Reproduction of the hallmark "AJPD40 (Rinhart n.29) present in Daguerreotype no. 3. Source: Chiesa and Gosio (2020). 3.5 x 9 mm.

Fig. 14. (a) General X-ray radiography of Daguerreotype no. 4. (b) Detail of the hallmark with Procedure 2. c) Detail of the hallmark with Procedure 1.

Fig. 15. Reproduction of the hallmark "HBEH40 (Rinhart n.20)" present in Daguerreotype no. 4. Source: Chiesa and Gosio (2020). 2 x 8 mm.

Fig. 16. (a) General X-ray radiography of Daguerreotype no. 5. (b) Location of the hallmark with Procedure 2. (c) Detail of the hallmark with Procedure 2.

Fig. 17. Reproduction of the "BALCDD" hallmark present on Daguerreotype no. 5. Source: Chiesa and Gosio (2020). 2.4 mm in length.

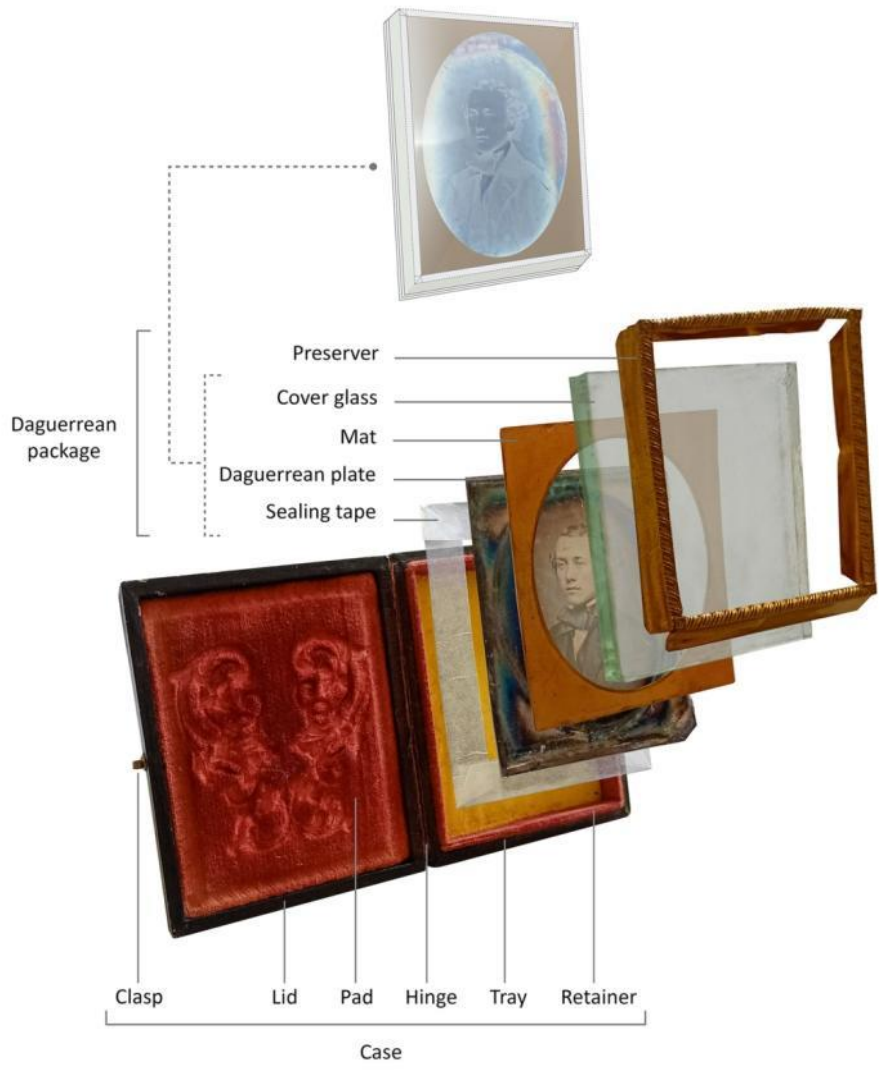

Figura 1

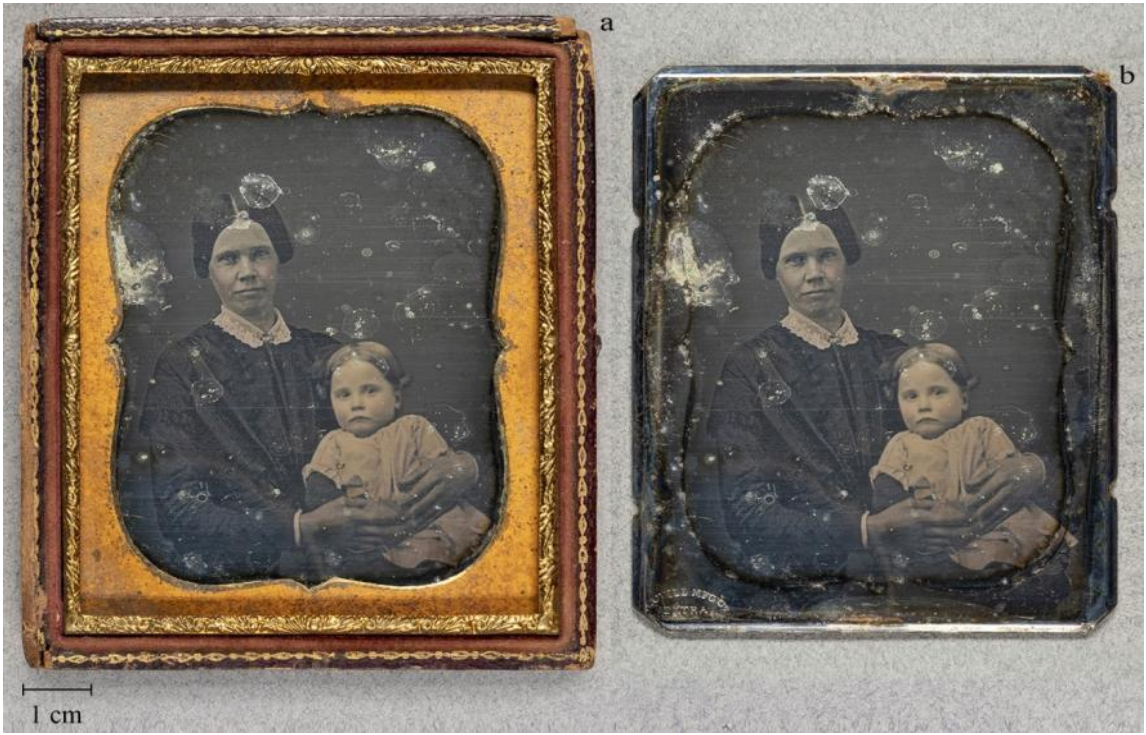

Figura 2

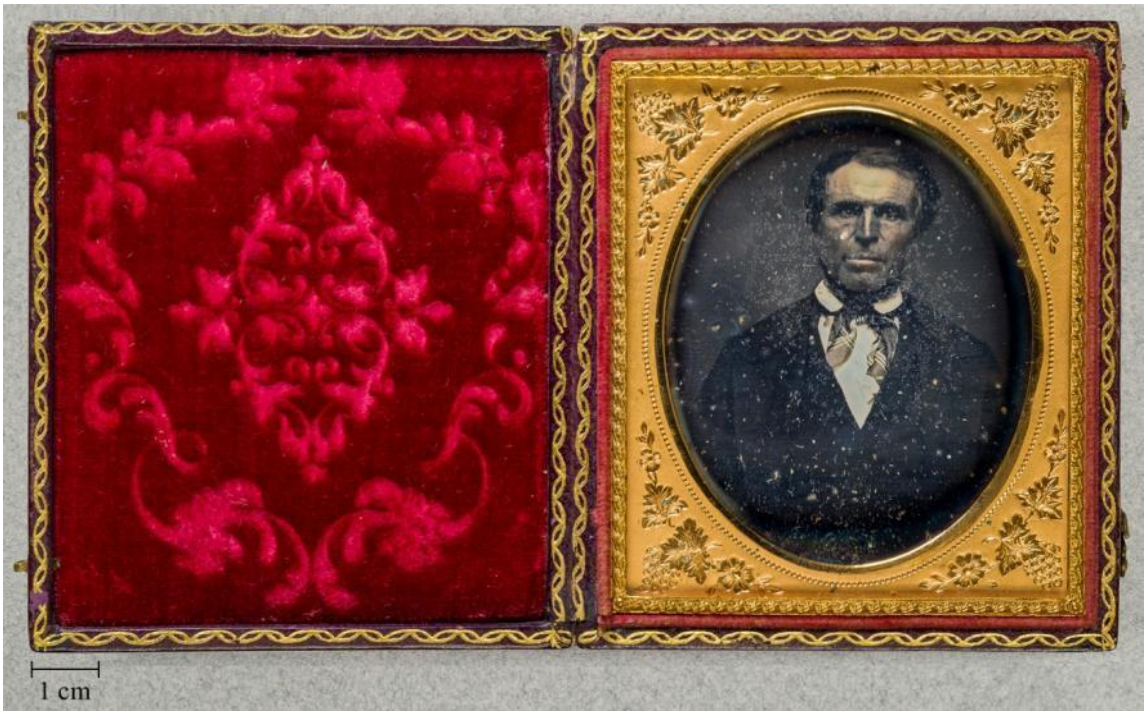

Figura 3

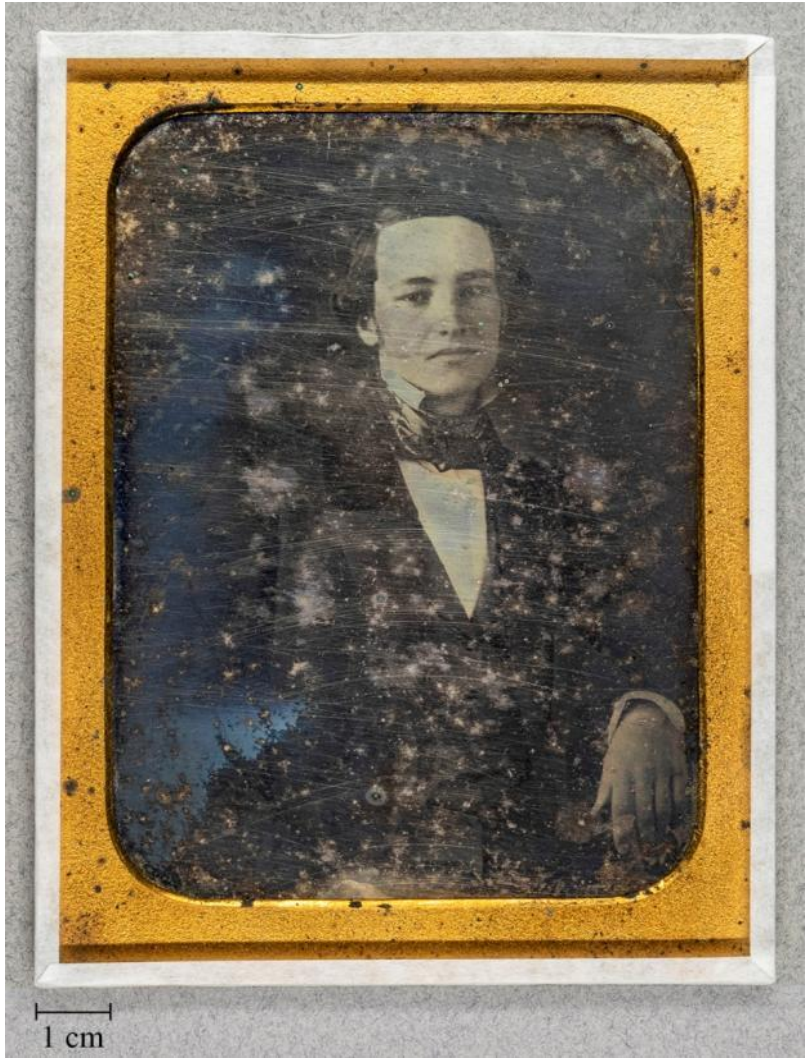

Figura 4

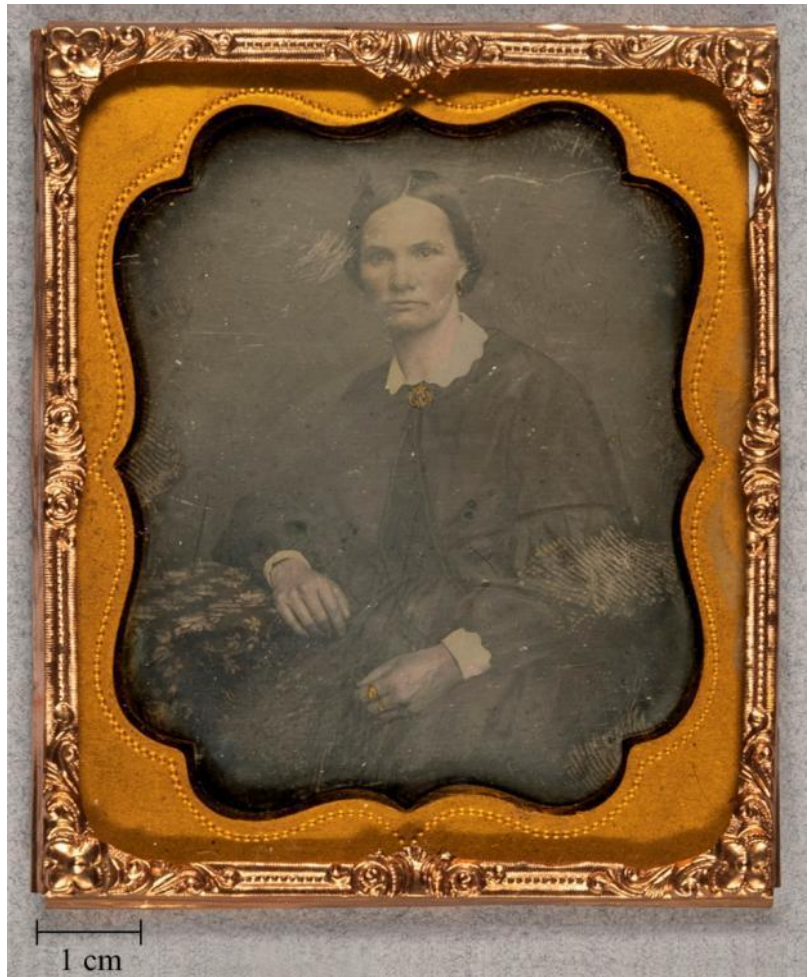

1 cm

Figura 5

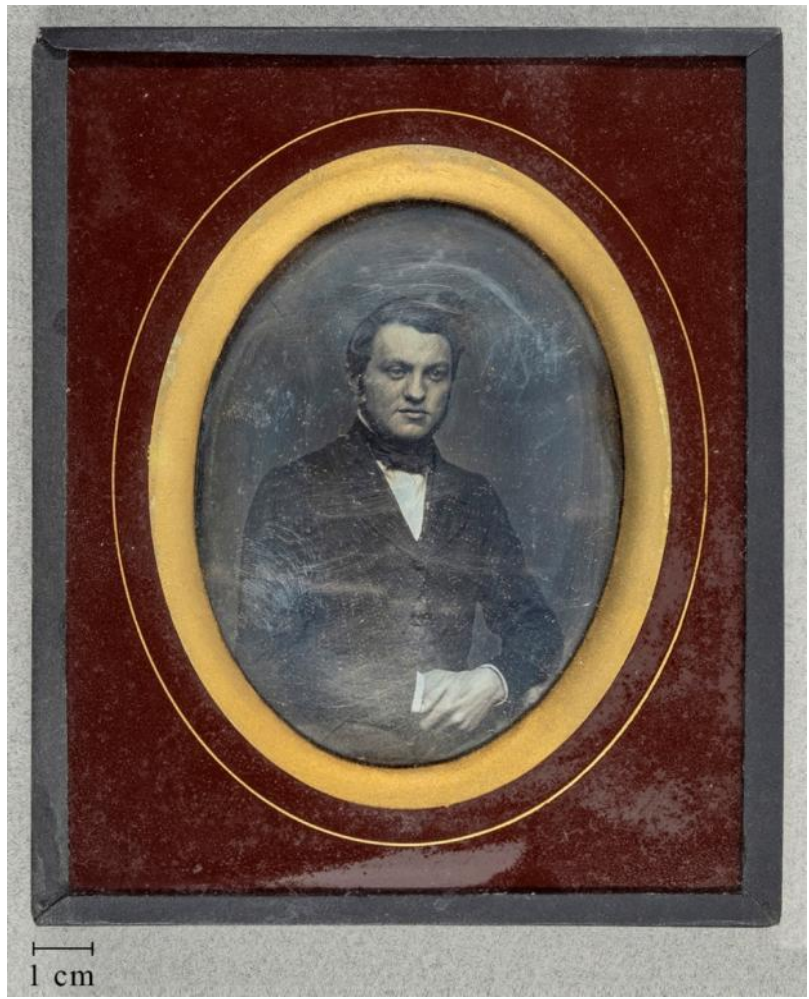

Figura 6

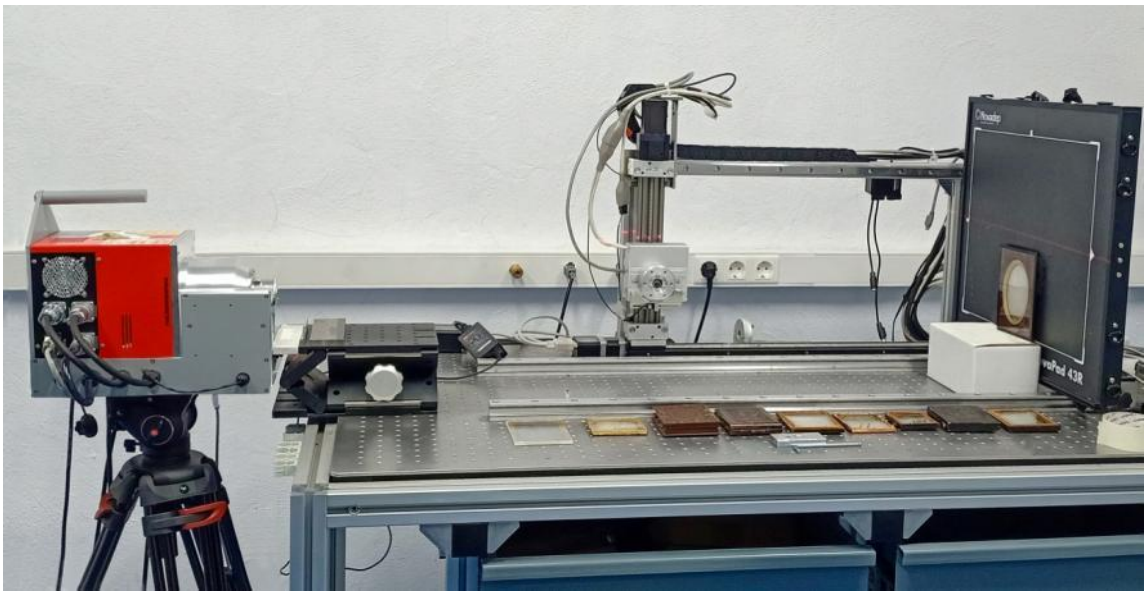

Figura 7

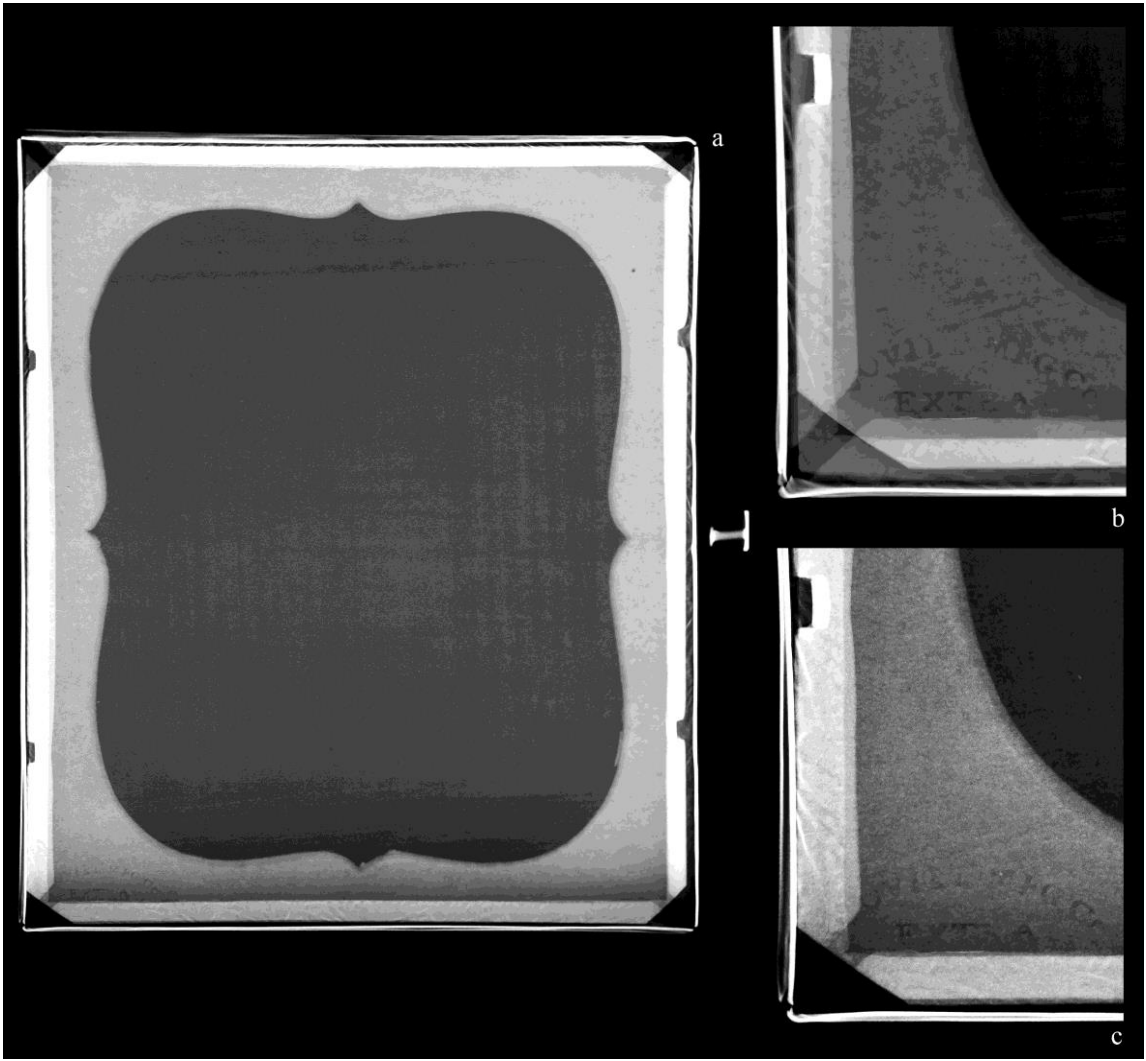

Figura 8

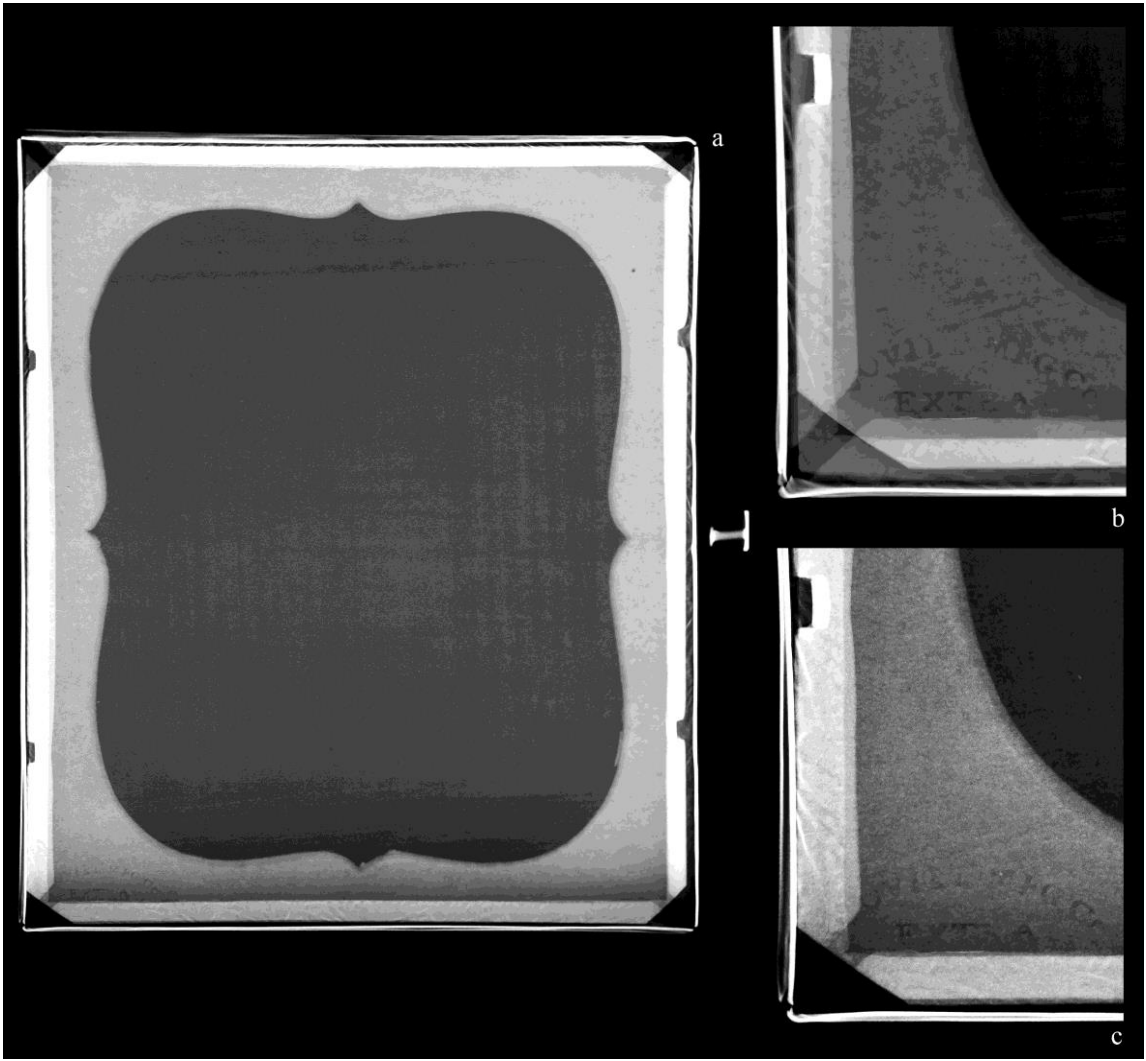

Figura 9

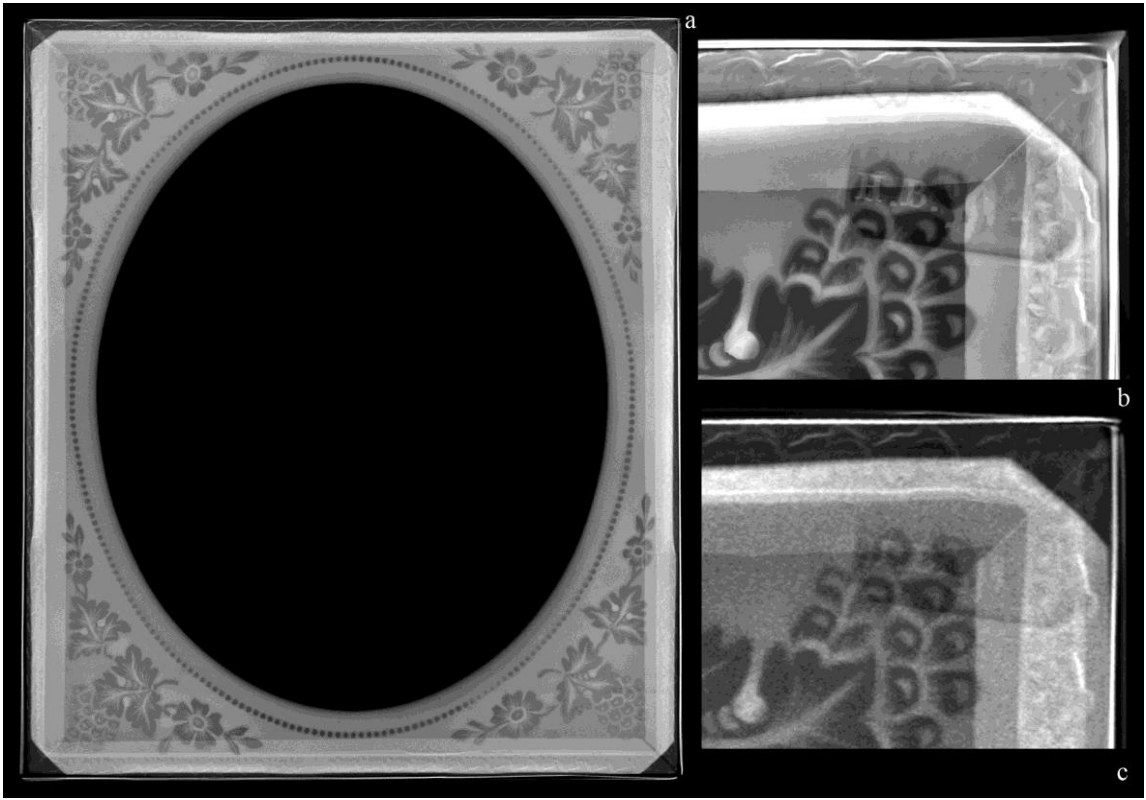

Figura 10

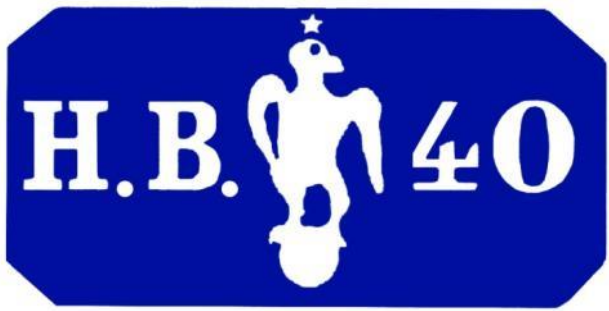

Figura 11

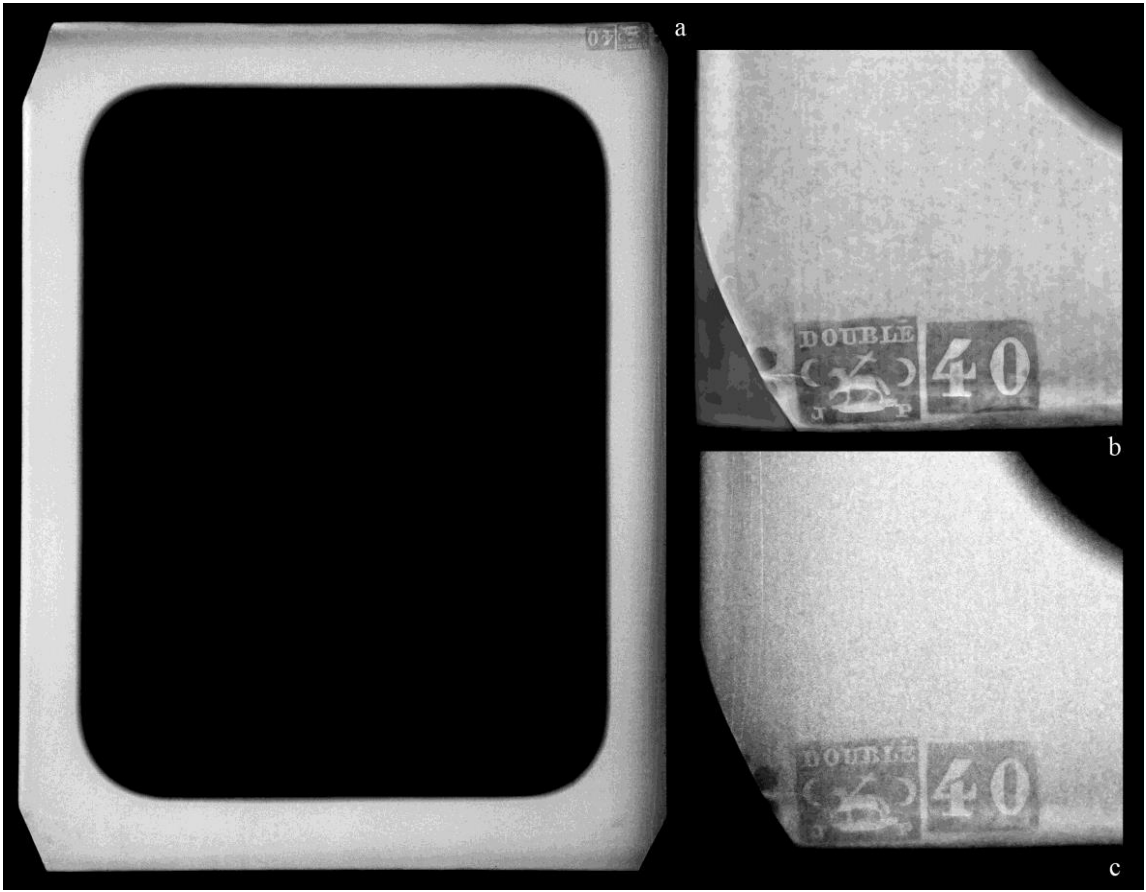

Figura 12

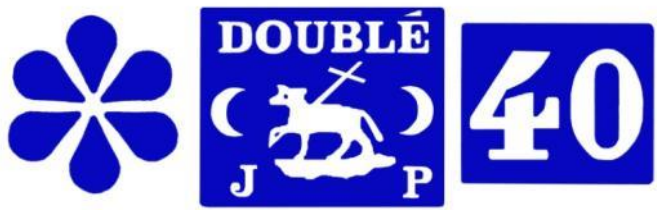

Figura 13

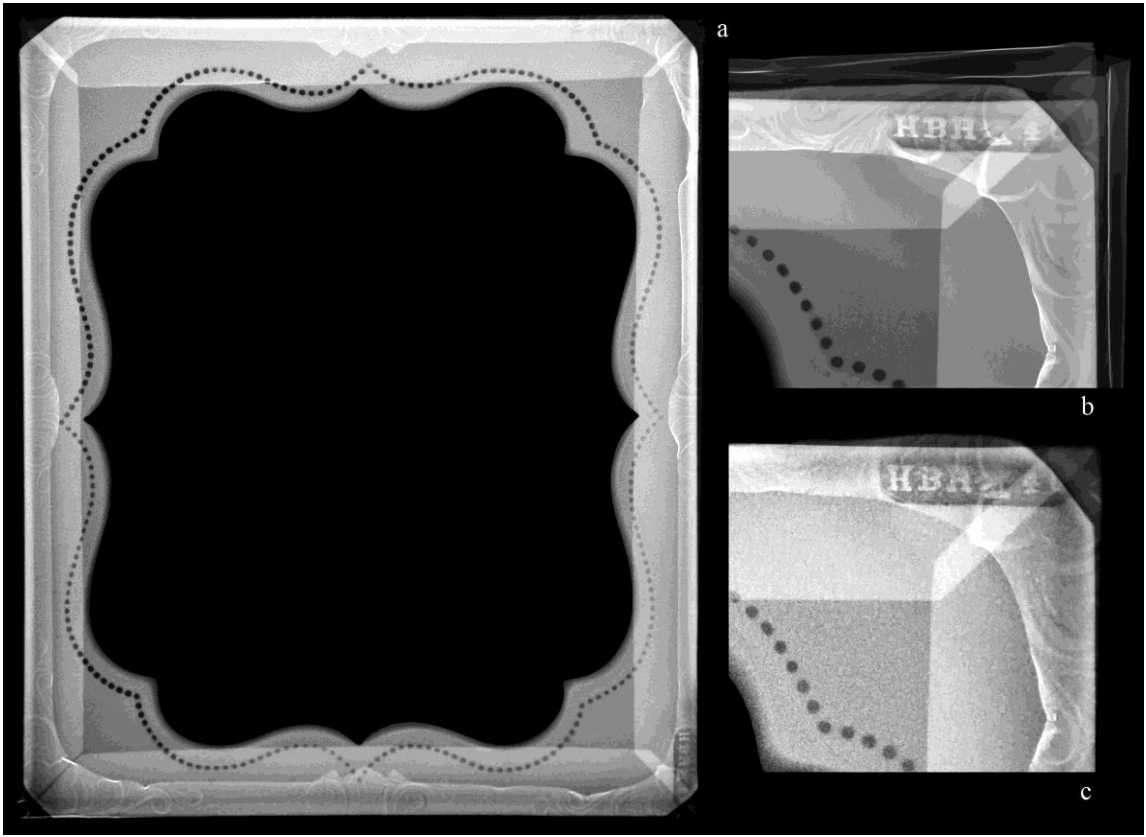

Figura 14

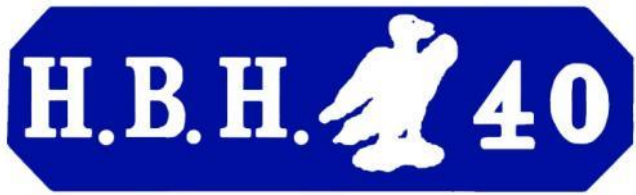

Figura 15

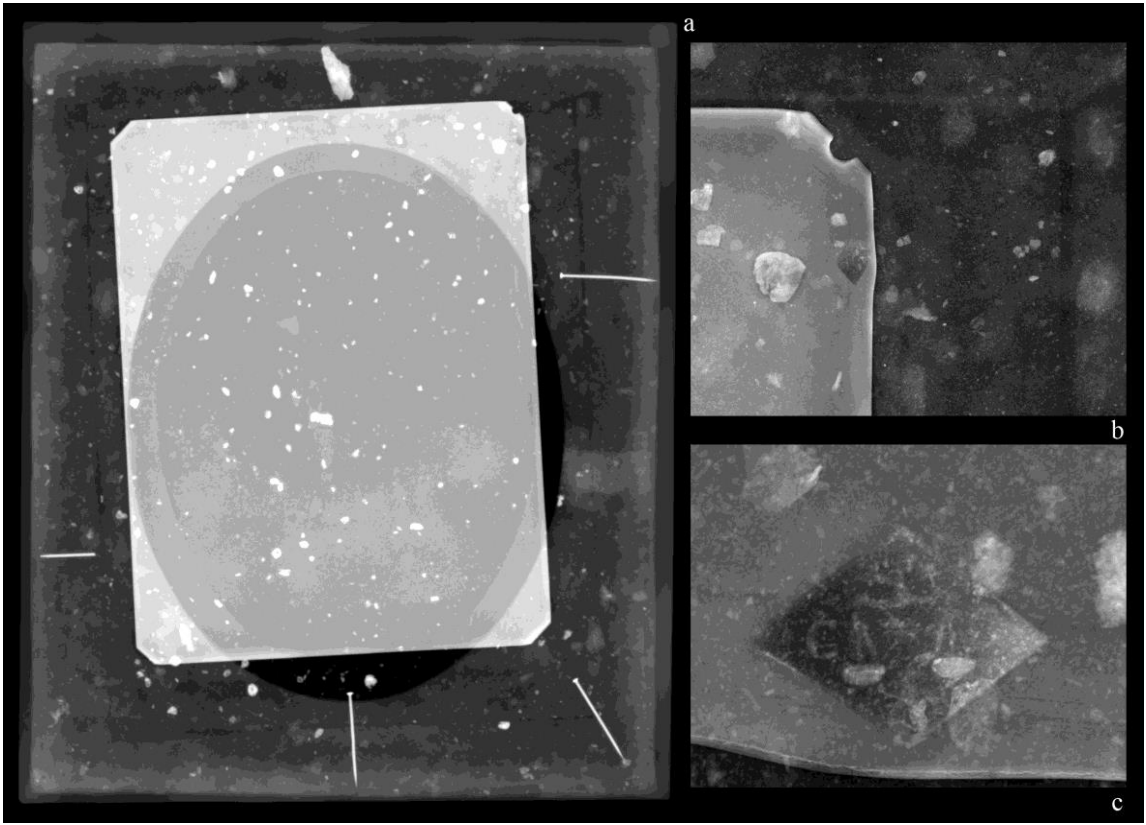

Figura 16

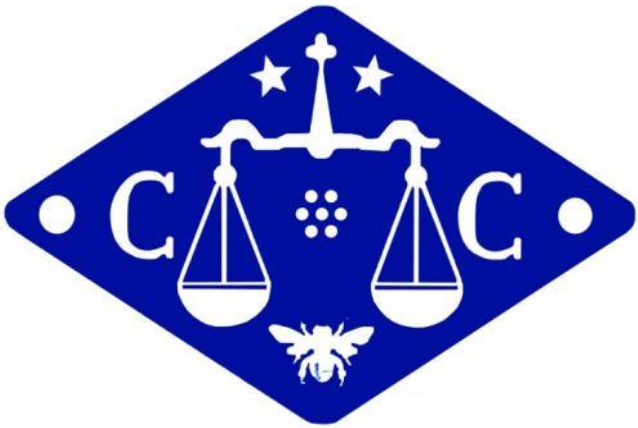

Figura 17